\documentclass[conference]{IEEEtran}

\usepackage{amsmath,amssymb,epsfig,cite,footnote,authblk}

\usepackage{color, hyperref}
\hypersetup{
    colorlinks,%
    citecolor=blue,%
    filecolor=blue,%
    linkcolor=blue,%
    urlcolor=blue
}
\definecolor{bluecolor}{rgb}{0,0.,1.}

\definecolor{redcolor}{rgb}{.7,0.,0.}

\newcommand{\es}[1]{\begin{equation}\begin{split}#1\end{split}\end{equation}}

\newcommand{\R}{\mathbb{R}}

\newcommand{\V}{\mathcal{V}}

\newcommand{\erf}{\mathrm{erf}}

\newcommand{\br}{\boldsymbol{r}}

\newcommand{\bX}{\boldsymbol{X}}
\newcommand{\dd}{\textrm{d}}

\newcommand{\Ck}{\color{black}}

\newcommand{\CB}{\color{blue}}

\begin{document}

\title{How does mobility affect the connectivity of interference-limited ad hoc networks?}
\author[1]{Pete Pratt}
\author[1]{Carl P. Dettmann}
\author[2]{Orestis Georgiou}
\affil[1]{School of Mathematics, University of Bristol, University Walk, Bristol, BS8 1TW, UK}
\affil[2]{Toshiba Telecommunications Research Laboratory, 32 Queens Square, Bristol, BS1 4ND, UK}
\maketitle


\begin{abstract}

One limiting factor to the performance of mobile ad hoc networks is the amount of interference that is experienced by each node. In this paper we will use the well established Random Waypoint Mobility Model (RWPM) to represent such a network of mobile devices, and show that the connectivity of a receiver at different parts of the network domain varies significantly. This is a result of a large portion of the nodes in the RWPM being located near the centre of the domain resulting in increased levels of interference between neighbouring devices. A non-trivial trade-off therefore exists between the spatial intensity of interfering signals and non-interfering (useful) ones. Using tools from stochastic geometry, we derive novel closed form expressions for the spatial distribution of nodes in a rectangle and the connection probability for an interference limited network indicating the impact an inhomogeneous distribution of nodes has on a network's performance. 
Results can therefore be used to analyse this trade-off and optimize network performance, for example through dynamic transmission schemes and adaptive routing protocols.

\end{abstract}

\section{Introduction \label{sec:intro}}

A mobile ad hoc network (MANET) is a self-configuring network of mobile devices making direct wireless links with each other rather than a central router; the topology of the network evolves with time as links are continually being made and broken. This network has various advantages over networks which have a fixed topology or a centralized structure including scalability (can continue to add more nodes), flexibility (can create temporary ad hoc networks anytime anywhere) and continuous reconfiguration which can enable the network to resolve any problems itself \cite{helen2014applications}. Such applications of these networks include environmental monitoring \cite{chong2003sensor}, disaster relief \cite{khan2012wireless} and military communications \cite{helen2014applications}. 

With the next generation of wireless communication (5G) in mind and as the number of personal devices such as mobile phones and tablets with access to the Internet continues to soar there has been a lot of work into the performance of MANETs. One of the first papers to highlight the potential advantages of an ad hoc network was by Grossgauler and Tse \cite{grossglauser2001mobility} who showed that by exploiting a network's mobility the increased throughput increases linearly, although this did not take into account the delay of the packets. As such this self configuring arrangement is seen to be a desirable feature for future mobile phone networks but would need to preserve coverage where mobile devices could disconnect and reconnect independent of location \cite{stanoevska2003impact}.

The aim of this paper is to look at modeling MANETs using the stochastic Random Waypoint Mobility Model (RWPM), which was proposed as one of the simplest models of MANETs mobility patterns \cite{camp2002survey}. 
This simplicity translates to mathematical tractability, and therefore has led to it being studied extensively in the literature as to model mobility in wireless networks \cite{gong2014interference,bettstetter2002spatial, hyytia2006spatial, kalogridis2009connection, singh2011comparative,kumar2011simulation}, and more recently in robotics \cite{rubenstein2012kilobot, ducatelle2014cooperative, kudelski2013robonetsim}.

The stationary distribution of RWPM allows us to more effectively model a MANETs performance  using the metrics of connection probability and mean network degree. 
Namely, we build upon previous work from Bettstetter and Hytti\"{a}, Lassila and Virtamo \cite{bettstetter2002spatial, hyytia2006spatial} where both considered the RWPM and investigated the probability of full connectivity on the unit disk, but give little indication as to the impact of interference would have on their results.
This further motivates our present work where we focus on the interference ﬁeld, and exemplify its dependence to the domain geometry \cite{haenggi2009stochastic}, \cite{banani2015analyzing}. Significantly, we contrast how the said MANET performance metrics vary for an inhomogeneous distribution of mobile nodes compared with the homogeneous case \cite{georgiou2015location}. 
Furthermore, we extend the interference models with inhomogeneous node distributions to domains other than circles \cite{srinivasa2007modeling}, \cite{dettmann2014more}. Employing tools from stochastic geometry, our analytical findings suggest that regions of high node density suffer from an intensified interference field thus hindering connectivity and coverage. 
Under the RWPM, such regions are typically found away from the domain borders, i.e. near the domain centre therefore strengthening the case for location aware MAC and routing protocols.

The main contributions of this paper are:
\begin{enumerate}
\item{We give an exact expression for the probability density function for the Random Waypoint Mobility model in a rectangle, and use the result to analyse the performance of the network.} 
\item{We give an expression for the interference in a network for a non-homogeneous density, and derive an analytic formula for the case where the receiving node is positioned at the centre of a circular domain. }
\item{We give an explicit expression for the mean degree of the network in the signal-to-noise regime and analyse it by comparing it with numerical calculations for the signal-to-interference-plus-noise regime.}
\end{enumerate}

The remainder of the paper is structured as follows: 
Sec. \ref{sec:RWPM} defines the Random Waypoint Mobility Model (RWPM) and gives an explicit expression for the probability density function (pdf) in a rectangular domain.
Sec. \ref{sec:Connectivity Analysis} defines the connectivity metric of interest and calculates the connection probability under the RWPM for different domain shapes and interference limits.
Sec. \ref{sec:mean_degree} utilises the expressions obtained for the connection probability to investigate the spatial density of successful transmissions.
Sec. \ref{sec:Conclusion} provides concluding remarks  and discusses potential areas of future work.

\section{Random Waypoint Mobility Model \label{sec:RWPM}}

\begin{figure}[t]
\centering
\includegraphics[scale=0.3, trim = {10 20 0 0 }, clip]{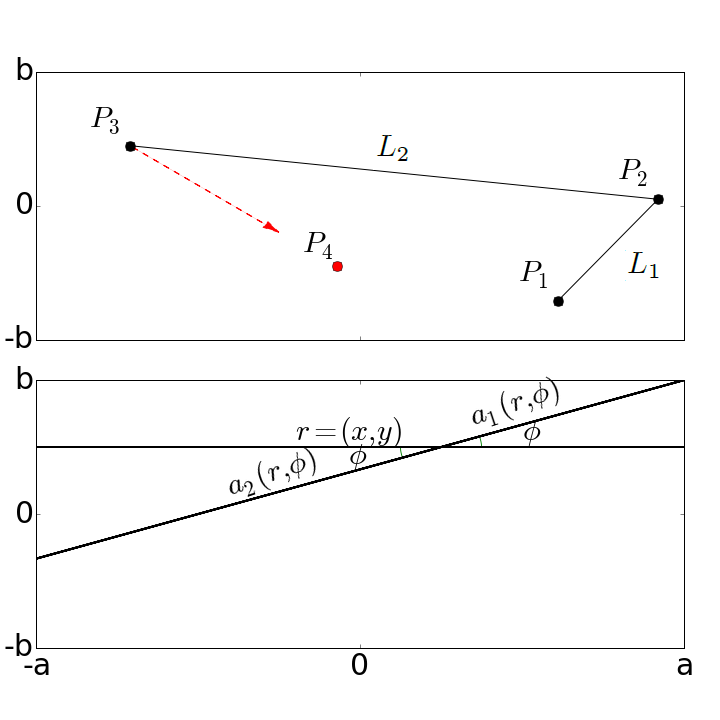}
\caption{\textit{Top}: A random realisation of the RWPM where a node on the domain $\mathcal{V}$ has travelled from its initial point $P_1$ to $P_2$, and then $P_3$, where it has chosen its next waypoint $P_4$ from a uniform distribution. 
\textit{Bottom}: A representation showing how $a_1$ and $a_2$ are defined (see eq \eqref{eq:l}), along with the angle $\phi$  at apoint $\br$ within $\mathcal{V}$.} 
\label{fig:fig1}
\end{figure}

\subsection{RWPM definition}
The RWPM assumes $N$ nodes randomly distributed inside a convex domain $\V\subseteq \R^2$ of area $V$ using a Binomial Point Process (BPP) of density $\rho=N/V$. 
At any given instance the node locations are given by $\br_i \in\V$ for $i=1,\ldots N$.
Each node moves independently from the other $N-1$ nodes so we can explain the process just by considering a single node (refer to Fig.~\ref{fig:fig1}). 
A node located at $P_1$ chooses a random waypoint $P_2$ uniformly inside $\V$ and travels to it in a straight line at a constant speed.
The speed is chosen from a uniform distribution of speeds $[v_{min}, v_{max}]$ where $ 0 < v_{min} \le k v_{min} = v_{max}$. 
This results in a sequence of waypoints defined as $\{P_1,P_2, \ldots\}$ and legs $\{L_{1}, L_2,\ldots \}$ which completely characterise the paths taken by a node. 
At each waypoint the node may pause for a ``think time" drawn from yet another uniform distribution $[0,T_{max}]$.
The RWPM reduces to a random walk model if the think time is zero.

\subsection{Spatial distribution of nodes under the RWPM}

\begin{figure}[t]
\centering
\includegraphics[scale=0.23]{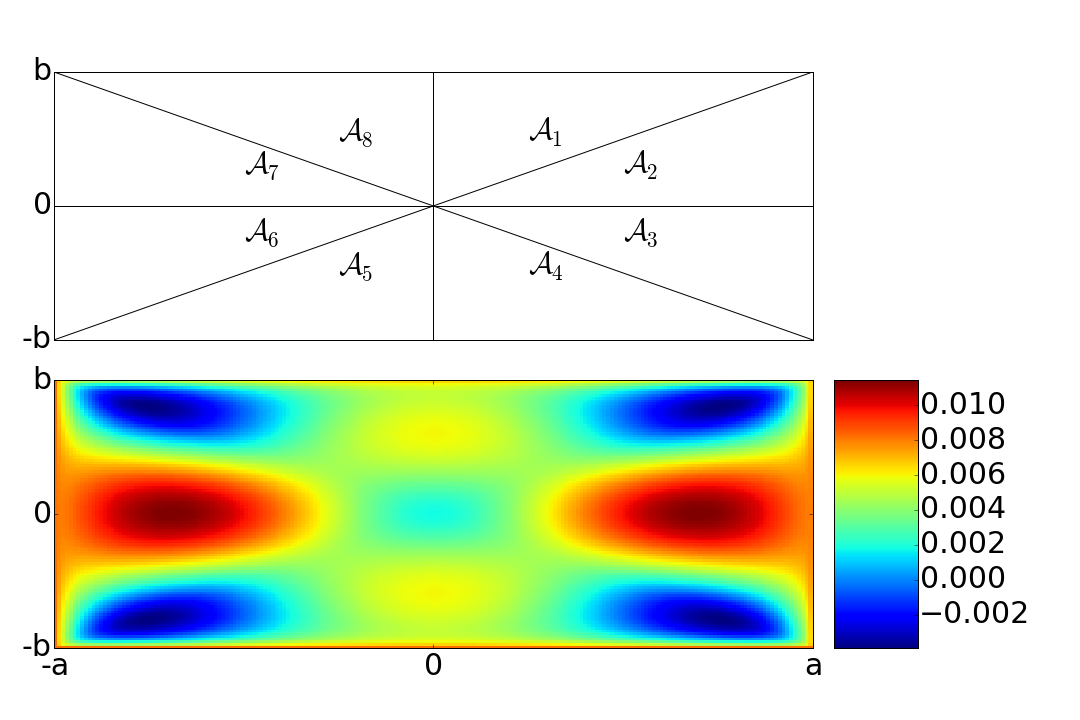}
\caption{\textit{Top}: The $8$ elementary segments of the rectangular domain in which $f_{\bX,m}(\br)$ is symmetrically identical.
\textit{Bottom}: The difference between the approximate distribution \eqref{eq:approx_pdf} and the exact one which follows from \eqref{eq:f1}.}
\label{fig:fig2}
\end{figure}

Defining $\wp_p\in[0,1]$ to denote the probability that a node is thinking (i.e. is not moving) we have that the RWPM nodes remain static if $\wp_p=1$, in which case the spatial distribution of nodes in $\V$ is uniform by definition.
This enables us to express the spatial distribution of nodes under the RWPM as a combination of mobile and static nodes probability density functions $f_{\bX,m}(\br)$ and $f_{\bX,p}(\br)$ respectively 
\es{
f_{\bX}(\br) = \wp_p f_{\bX,p}(\br) + (1-\wp_p)f_{\bX,m}(\br)
\label{eq:gen_pdf}
,}
such that $f_{\bX,p}(\br)=1/V$ and  $\int_\V f_{\bX}(\br) \dd \br =1$ and where
\es{
\wp_p =\frac{\mathbb{E} [ T_p ]}{\mathbb{E} [ T_p ] + \mathbb{E} [ T ]},
\label{eq:full_pdf}
}
as shown in \cite{bettstetter2004stochastic}, where $\mathbb{E}[T_p ]$ is the expected pause time and $\mathbb{E} [ T ]$ is the mean time taken for a single leg. 
Using the fact that the speed is taken from a uniform distribution and that $v_{max} = k v_{min}$, the expected time for each leg is given by
\es{
\mathbb{E} [T ] = \frac{\ln(k)}{k - 1}   \frac{\overline{l}}{v_{min}}
,}
where $\overline{l}$ is the mean leg length and is exactly given by \cite{crescenzi2010spatial}
\es{
\overline{l}=\frac{1}{V^2} \int_\V  \int_0^\pi a_1 a_2 (a_1+a_2) \dd \phi \dd \br
\label{eq:l}
,}
for any convex domain $\V$, where the lengths $a_1$ and $a_2$ depend on the position $\br\in \V$ of the node as demonstrated in the lower panel of Fig.~\ref{fig:fig1} for a rectangular domain.
The values of $a_1, a_2$ are defined by choosing a point $\br\in\V$ and drawing a straight line through it.
The length of the line to the right is $a_1$ and to the left $a_2$.

Similarly, the exact spatial distribution of nodes in any convex shape $\V$ can be found by calculating \cite{hyytia2006spatial}
\es{
f_{\bX,m}(\br) = \frac{1}{\overline{l}V^2}\int_0^{\pi} a_1a_2(a_1 + a_2) d\phi
\label{eq:Integral}
.}

It is interesting to note that the distribution of nodes in the RWPM, given by \eqref{eq:Integral} is proportional  to that for the betweenness centrality measure where a node is said to have a high betweenness if it is frequently used to transmit data between two different nodes along the shortest path \cite{giles2014betweenness}. This allows for some intuition as to what the spatial distribution of nodes in the RWPM will be like since in \cite{giles2014betweenness} it was shown the nodes located at the centre had a high betweenness value so analogously we should expect the distribution of nodes to be highest in the centre also.

\subsubsection{Rectangular domain}

Writing $a_1$ and $a_2$ as functions of position $\br$ and $\phi\in[0,\pi]$ one can calculate that for a rectangle of sides $a$ and $b$  \cite{bettstetter2004stochastic}
\es{
\overline{l} \!= \!\frac{d}{3} \!+\! \frac{a^2}{6b} \ln(\frac{{{d}} + b}{a}) - \frac{b^2}{6a}\ln(\frac{{d}-a}{b}) + \frac{{a^3} - {d}^3}{15{b^2}} + \frac{{b^3} - {{d}^3}}{15{a^2}}
\label{eq:mean_length_rect}
,}
where $a \ge b$ and $d = 2\sqrt{a^2 + b^2}$.
Note that when $a_1$ and $a_2$ meet the corners of the rectangle the integrands above are not smooth functions thus requiring that the integrals be performed within the elementary cells $\mathcal{A}_i$  for $i=1,\ldots 8$ as shown in Fig.~\ref{fig:fig2}

We give here for the first time the exact spatial distribution of mobile nodes under the RWPM within $\mathcal{A}_1$ elementary cell of a rectangular domain, where the full expression follows through symmetry.
\es{
f_1(x,y) &= \frac{1}{16a^2b^2\overline{l}} \Biggl\{\frac{d_1(b-y)(a+x)(a-2x)}{(a-x)}\\&+ \frac{d_4(a-x)(a+2x)(b-y)}{(a+x)}\\&+ \left[d_3(a-x) + d_2(a+x)\right]\left[\frac{(b-y)(b+2y)}{(b+y)}\right] \\&+  2x(b-y)^2\ln\left|\frac{a-x}{a+x}\right| + 4ax(b-y)\ln\left|\frac{b-y}{b+y}\right|\\&+ c_1\left[\ln\left|\frac{b-y+d_1}{a-x}\right|+ \ln\left|\frac{y-d_4-b}{a+x}\right|\right] \\&+ c_2\ln\left|\frac{d_3+a-x}{d_2-a-x}\right| + c_3\ln\left|\frac{d_2+b+y}{d_1+b-y}\right|\\&+ c_4\ln\left|\frac{d_2-a-x}{d_1+x-a}\right| - c_5\ln\left|\frac{y- d_4 - b}{-d_3 - b- y}\right| \\&+  c_6\ln\left|\frac{d_4 + a +x}{d_3+a-x}\right|\Biggl\}
\label{eq:f1}
,}
where $c_1 = a(a-x)(a+x),\; c_2 = b(b-y)(b+y),\; c_3 = (a+x)(b-y)^2,\; c_4 = (a+x)^2(b-y),\; c_5 = (x-a)(b-y)^2 \; \text{and}\; c_6 = (x-a)^2(b-y)$, $d_1 =\sqrt{(a-x)^2 + (b-y)^2}$, $d_2 = \sqrt{(a+x)^2 + (b+y)^2}$, $d_3 = \sqrt{(a-x)^2 + (b+y)^2} $ and $d_4 = \sqrt{(a+x)^2 + (b-y)^2}$.

Several approximations to the above have been reported in the literature the simplest one being \cite{bettstetter2002spatial} 
\es{
\hat{f}_{\boxdot}(x,y)&= \frac{9}{16a^3b^3}(x^2 - a^2)(y^2 - b^2) 
\label{eq:approx_pdf}
,}
by assuming two linearly independent processes in the $x$ and $y$ directions.
The $\boxdot$ is used to indicate the rectangular domain.
Equation \eqref{eq:approx_pdf} deviates from the exact solution as seen in the lower panel of Fig.~\ref{fig:fig2}.
Hyyti\"{a} et al \cite{hyytia2006spatial} give a more accurate polynomial approximation than \eqref{eq:approx_pdf} however, unlike the exact expression \eqref{eq:f1}, all known approximations are smooth everywhere whereas the exact solution is piecewise continuous. 

Later in this paper we will use the approximation given by \eqref{eq:approx_pdf} due to its simplicity and thus ease in calculating complicated connectivity integrals.
We jsutify our use of the approximations as we hope to give closed form expressions where possible and argue that the results will not change qualitatively.

\subsubsection{Circular domain}

A circular domain is much easier to analyse than the rectangle as there are no discontinuities due to corners.
Therefore, by using  \eqref{eq:Integral}, it can be shown that 
\es{
f_{\bigodot}(r,\theta)&= \frac{2(R^2-r^2)}{\overline{l} V^2}\int_0^{\pi} \sqrt{R^2 - r^2\cos^2(\phi)}\: d\phi
\label{eq:f_disk_exact}
,}
in polar coordinates $(r,\theta)$, and $\overline{l}$ can be found by integrating \eqref{eq:f_disk_exact} over  $\mathcal{V}$, which can be approximated to \cite{bettstetter2002spatial} 
\es{
\hat{f}_{\bigodot}(r) & =  \frac{2}{\pi R^2} \left(1 - \left(\frac{r}{R}\right)^2\right)
\label{eq:disk_pdf}
,}
where $R$ is the radius of the circular domain.


\section{Connectivity Analysis \label{sec:Connectivity Analysis}}

Given $N$ nodes which move under the RWPM within some convex domain $\mathcal{V}$, in this section we will investigate the connectivity of the resulting network.
We first present the information theoretic model \cite{tse2005fundamentals} that we adopt and later define the observables of interest.

\subsection{System Model}

The attenuation in the wireless channel of any ad hoc network affects the overall connectivity and capacity of that network \cite{dousse2004connectivity}.
As a result we introduce the path loss function $g(d_{ij})$ describing how the power of a propagating signal decays with the distance $d_{ij}=|\br_i - \br_j|$ between two nodes.
We assume that the function $g(d_{ij})$ is only concerned with the long term average of the signal to noise ratio (SNR) at the receiver which behaves like SNR$_{ij}$ $\propto d_{ij}^{-\eta}$ and so we define the path loss function $g(d_{ij})$ as 
\es{
g(d_{ij}) &= \frac{1}{\epsilon + d_{ij}^{\eta}}\:\:\text{, $\epsilon \ge 0$}
\label{eq:path_loss}
,}
where $\eta$ is the path loss exponent and $\epsilon$ is chosen to be non-zero so the path loss function is non-singular. 
For free space propagation it is common to take $\eta = 2$ and for more cluttered environments $\eta > 2$, typically taking values in the range $[2,6]$. 

We now turn to the main connectivity metric the Signal to interference plus noise ratio (SINR) defined as
\es{
\text{SINR}_{ij} &= \frac{\mathcal{P}|h_{ij}|^2g(d_{ij})}{\mathcal{N} + \gamma \mathcal{I}_j}\\
\mathcal{I}_j &= \sum _{k \neq i} \mathcal{P}|h_{kj}|^2g(d_{kj})
\label{eq:SINR}
,}
where $\mathcal{I}_j$ is the interference received at node $j$,
$\mathcal{P}$ is the transmit power (equal for all nodes), 
$\mathcal{N}$ is the noise power within the system,
and $|h_{ij}|^2$ is the channel gain between nodes $i$ and $j$ and will be modelled as an exponential random variable with mean one (assuming Rayleigh fading).
In \eqref{eq:SINR} $\gamma$ is used to quantify the amount of interference in the system and can take values between zero and one. 
In the case where $\gamma = 0$ there is no interference resulting in all $k \neq i$ interfering devices transmitting on a different channel \cite{yang2012connectivity}. Conversely $\gamma = 1$ refers to the case when all transmissions occur in the same channel.

In the RWPM (and other mobility models) the SINR between two nodes depends on the location of the receiver, the underlying network topology and the spatial distribution of the network defining the interference field experienced by the receiver. 
All these can significantly affect the performance of the network.
Intuitively we can see that dense regions will have more interference than sparse ones.
On the other hand, nodes in sparsely populated regions of the network domain have less neighbours to connect to compared to dense regions.
This trade-off is already non-trivial in the case of a uniform distribution of interfering nodes \cite{georgiou2015location}.
In this paper we will study for the first time this trade-off in the non-uniform distribution \eqref{eq:gen_pdf} generated by the RWPM.

\subsection{Connection Probability} 

Let us first consider an interference limited network where node $i$ sends a signal to node $j$ and we assume that all $k \neq i$ nodes are interfering with that transmission and ask what is the probability $H_{ij}$ that node $i$ can successfully transmit data to node $j$, i.e. the complement of the outage probability given some SINR threshold $q$
\es{
H_{ij} = \mathbb{P}[\text{SINR}_{ij} > q] = \mathbb{P}\left[|h_{ij}|^2 \ge \frac{q(\mathcal{N} + \gamma \mathcal{I}_j)}{\mathcal{P}g(d_{ij})}\right]
\label{eq:connect_prob}
}
By conditioning on $\mathcal{I}_j$ and using the fact $|h_{ij}|^2 \sim  \text{exp}(1)$ and that $|h_{kj}|^2$ are i.i.d random variables, \eqref{eq:connect_prob} can be rewritten as,
\es{
H_{ij} &=\mathbb{E}_{\mathcal{I}_j}\Biggl[\mathbb{P}\left[|h_{ij}|^2 \ge \frac{q(\mathcal{N} + \gamma \mathcal{I}_j)}{\mathcal{P}g(d_{ij})}\biggl|\mathcal{I}_j\right]\Biggl]\\
&= \mathbb{E}_{\mathcal{I}_j}\Biggl[\exp \left(-\frac{q(\mathcal{N} + \gamma \mathcal{I}_j)}{\mathcal{P}g(d_{ij})}\right)\Biggl]\\
&= e^{-\frac{q\mathcal{N}}{\mathcal{P}g(d_{ij})}}\mathcal{L}_{\mathcal{I}_j}\left(\frac{q\gamma}{\mathcal{P}g(d_{ij})}\right)\\
\label{eq:connect_prob_2}
}
where $\mathcal{L}_{\mathcal{I}_j}(s)$ is the Laplace transform of the random variable $\mathcal{I}_j$ evaluated at $s = \frac{q\gamma}{\mathcal{P}g(d_ij)}$ conditioned on the locations of the transmitting and receiving nodes.
Following \cite{georgiou2015location} we have
\es{
\mathcal{L}_{\mathcal{I}_j}(s) &= \mathbb{E}_{\mathcal{I}_j}\left[e^{-\frac{s\mathcal{I}_j}{\mathcal{P}}}\right] = \mathbb{E}_{|h_{kj}|^2, d_{kj}}\left[e^{-s\sum_{k \neq i}  |h_{kj}|^2g(d_{kj})}\right]\\
&= \mathbb{E}_{d_{kj}}\left[\prod_{k \neq i}^{N - 1}  \frac{1}{1+s g(d_{kj})} \right]
\label{eq:Laplace}
.}
Invoking the probability generating functional of a general inhomogeneous Poisson point process (PPP) $\Xi$ in $\R^2$ with intensity function $\lambda(\xi)$ given by 
\es{\mathbb{E}\left[\prod_{\xi \in \Xi} f(\xi)\right] = \exp\left(-\int_{\mathbb{R}^2} (1 - f(\xi))\lambda(\xi)\dd \xi\right)
}
we can see that when $N\gg 1$ we can approximate the BPP by a PPP such that $\lambda(\xi)\approx N f_{\bX}(\br)$ therefore arriving at
\es{
\mathcal{L}_{\mathcal{I}_j} (s)
&\approx \exp \left(-N\int_{\mathcal{V}} f_{\bX}(\br_k) \frac{sg(d_{kj})}{1 + sg(d_{kj})} \dd\br_k\right)
\label{eq:Laplace2}
.}
\textit{Remark 1:} Equation \eqref{eq:Laplace2} is the main result of this paper as this can be computed numerically with the assistance of \eqref{eq:Integral} for any domain $\V$ in which nodes move under the RWPM. Significantly, we note that the connection probability $H_{ij}$ of a receiver at $\br_{j}$ given by \eqref{eq:connect_prob_2} depends exponentially on the node distribution $f_{\boldsymbol{X}}$ which is itself dependent on the mobility model.
We now proceed to consider specific cases.

\subsubsection{Circular domain}

Consider a receiver positioned at the centre of the circular domain i.e. $\br_j= 0$.
Using polar coordinates and the approximation given in \eqref{eq:disk_pdf} with $\rho = \frac{N}{V}$ we calculate
\es{
\mathcal{L}_{\mathcal{I}_j}(s) &= \exp\biggl\{-\frac{\wp_p \rho s\pi R^2}{\epsilon + s} {}_2F_1\left(1, \frac{2}{\eta},\frac{2}{\eta} + 1, \frac{-R^{\eta}}{\epsilon + s}\right)\\
&- \frac{(1-\wp_p)\pi R^2s \rho}{s + \epsilon}\biggl[2{}_2F_1\left(1, \frac{2}{\eta}, \frac{2+\eta}{\eta}, -\frac{R^{\eta}}{s+\epsilon}\right)\\
&-  {}_2F_1\left(1, \frac{4}{\eta}, \frac{4}{\eta} + 1,-\frac{R^{\eta}}{s + \epsilon}\right)\biggl]\biggl\}
\label{eq:SINR_disk}
.}
For free space propagation $\eta = 2$ we can simplify \eqref{eq:SINR_disk} to
\es{
\mathcal{L}_{\mathcal{I}_j}(s) &= \exp\biggl\{-\wp_p s \pi \rho \ln\left(\frac{R^2}{s+\epsilon} + 1\right)\\
&+ \frac{2 \rho \pi s (1-\wp_p)}{R^2}\left[R^2 - (s+\epsilon + R^2)\ln\left(\!\frac{R^2}{s+\epsilon} + 1\right)\!\right]\biggl\}
\label{eq:SINR_disk_eta_2}
.}

Similarly, for $\eta=4$ we can simplify \eqref{eq:SINR_disk} to
\es{
\mathcal{L}_{\mathcal{I}_j}(s) &= \exp\biggl\{-\frac{\wp_ps\pi \rho}{\sqrt{s+\epsilon}} \arctan \left(\frac{R^2}{\sqrt{s+\epsilon}}\right)\\
&+\frac{\rho \pi s (1 - \wp_p)}{\sqrt{s + \epsilon}}\biggl[\frac{\sqrt{s+\epsilon}}{R^2}\ln\left(\!\frac{R^4}{s+\epsilon} + 1\right)\!\\
&-2\arctan\left(\frac{R^2}{\sqrt{s+\epsilon}}\right)\biggl]\biggl\}
\label{eq:SINR_disk_eta_4}
.}

\subsubsection{Comparison with other models}

\begin{figure}[t]
\centering
\includegraphics[scale=0.2]{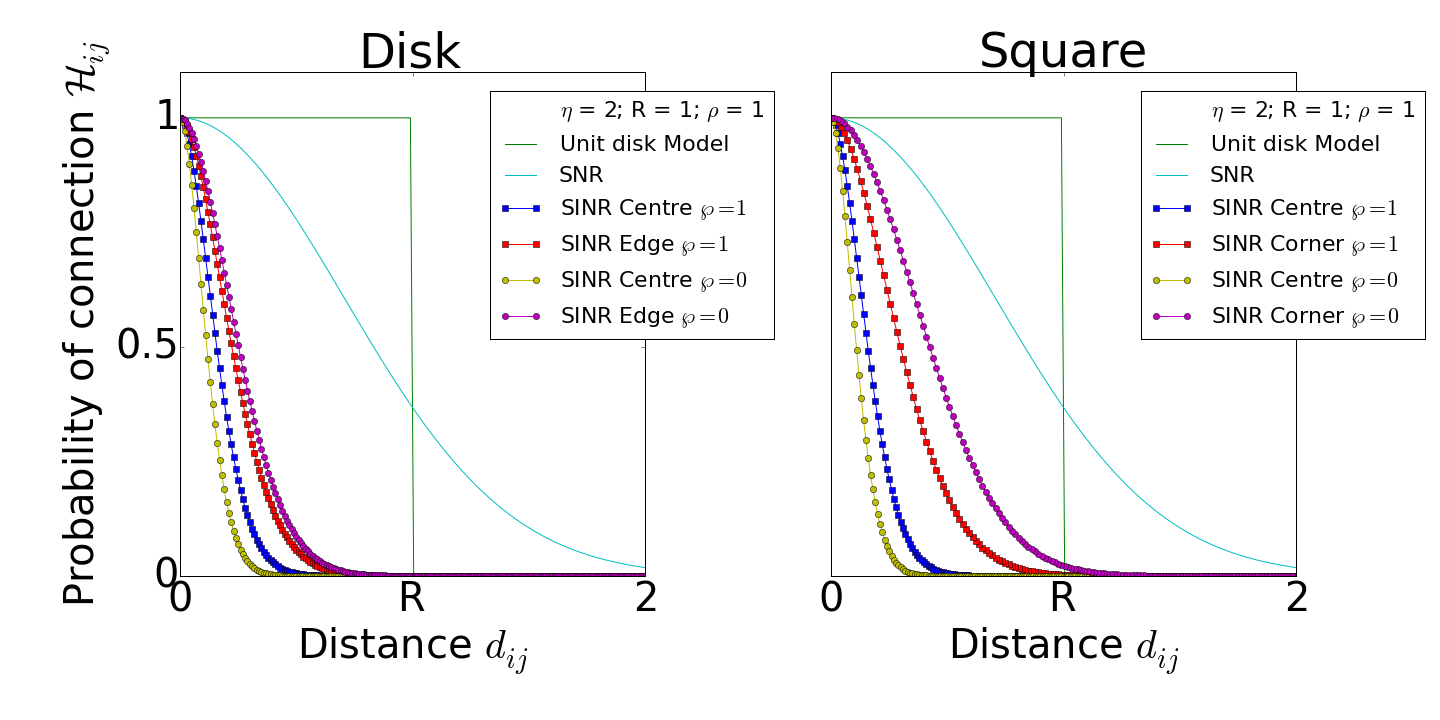}
\caption{A comparison between the connectivity from the centre to the edge in all three cases where the volume of the domain is given by $4ab$, with $a=b=10$,  $R = \frac{2a}{\sqrt{\pi}}$ and $\rho = NV$.}
\label{fig:fig3}
\end{figure}

\begin{figure}[t]
\centering
\includegraphics[scale=0.5, trim = 500 10 500 10]{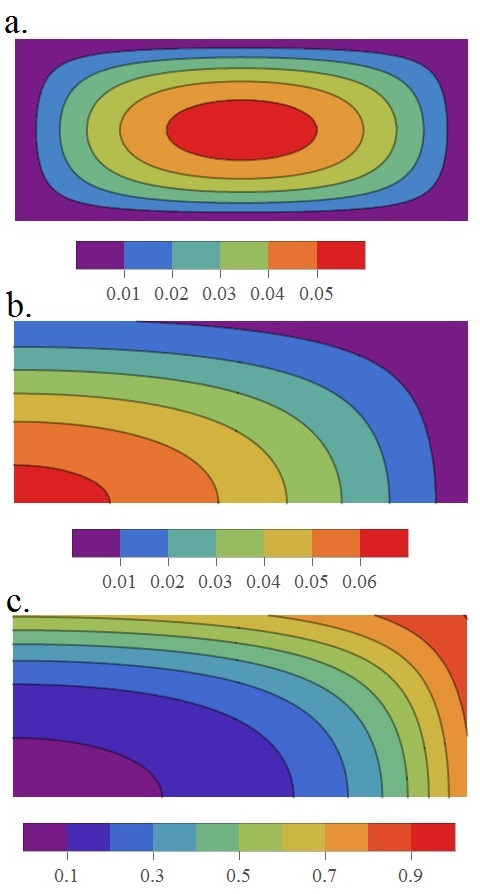}
\caption{a. The spatial node distribution  under the RWPM in a rectangular domain $\V$ 
b. The interference field in the top right quadrant of $\V$ as calculated through the integral of equation \eqref{eq:Laplace2}.
c. The connection probability $H_{ij}(1/2)$ of a receiver in the top right quadrant of the $\V$ as calculated through equation \eqref{eq:connect_prob_2} for $N=40$.
Parameters used: $\mathcal{P}=\mathcal{N}=q=\gamma=1$, $a=5$, $b=2$, $\eta=4$, and $\epsilon=\wp_p=0$.}
\label{fig:fig3b}
\end{figure}

In this subsection we compare the derived performance metric, the connection probability between two nodes $H_{ij}$, with a number of other models and receiver locations.
The comparison is shown in Fig.~\ref{fig:fig3} and is performed in both a Circular (Disk) domain (left panel) and a rectangular domain (right panel) using $q= \mathcal{P}= \mathcal{N} = \rho= 1$ and $\epsilon=0$.
The connection probability is plotted as function of transmitter-receiver distances $d_{ij}$ for different receiver positions: centre, edge, and corner of the domain $\V$ for $\wp_p=1$ (uniform distribution) and $\wp_p=0$.
Also plotted in Fig.~\ref{fig:fig3} is the case of $\gamma=0$ (i.e. no interference) referred to in the key as SNR, and also the case where $(\gamma,\eta)= (0,\infty)$ which interestingly corresponds to a deterministic unit disk model where nodes connect (or not) if they are within a unit distance of each other.
We notice that interference significantly deteriorates the connection probability.
Moreover, we notice that the receiver nodes near the border of the domain are more likely to be connected than  receiver nodes at the domain centre.
Interestingly, random waypoint mobility (i.e. $\wp_p<1$) improves the connection probability near the domain border, and worsens it near the domain centre.
This is not surprising as the density of interfering nodes near the centre is much higher than near the domain border (c.f. \eqref{eq:approx_pdf}).
This effect seems to be even more dominant in the rectangular domain. For instance, $H_{ij}(0.5) \approx 0$ for a node near the centre of the rectangular domain whilst , $H_{ij}(0.5) \approx 0.5$ for a node located near the domain corner; a huge difference in performance. 

The effect of interference on connectivity is further illustrated by  Fig.~\ref{fig:fig3b} where we observe in a rectangular domain with $N = 40$ that $H_{ij} \approx 0.1$ for a node located at the centre where as at the corner $H_{ij} \approx  0.8$. Clearly this is a direct consequence of the interference field as shown in Fig.~\ref{fig:fig3b}b. The effect a rectangular domain has on the connectivity is further highlighted by seeing that the side with the shortest length has a lower distribution of nodes and thus the impact of interference is less. 

Note that similar observations are expected to hold for the average achievable rate given by $\mathbb{E}[ \ln (1+\mathrm{SINR}_{ij}) ]$ (see \cite{georgiou2015location}).

\section{Spatial density of Successful Transmissions \label{sec:mean_degree}}

We now turn to the spatial density of transmissions that can be successfully received by a receiver located at $\br_j$ given by
\es{
\mu_j(\br_j) = (N-1) \int_\V f_{\bX}(\br_i) H_{ij} \dd \br_i
\label{eq:mu}
.}
\begin{figure}[t]
\centering
\includegraphics[scale=0.2, trim = {80 45 0 40}, clip]{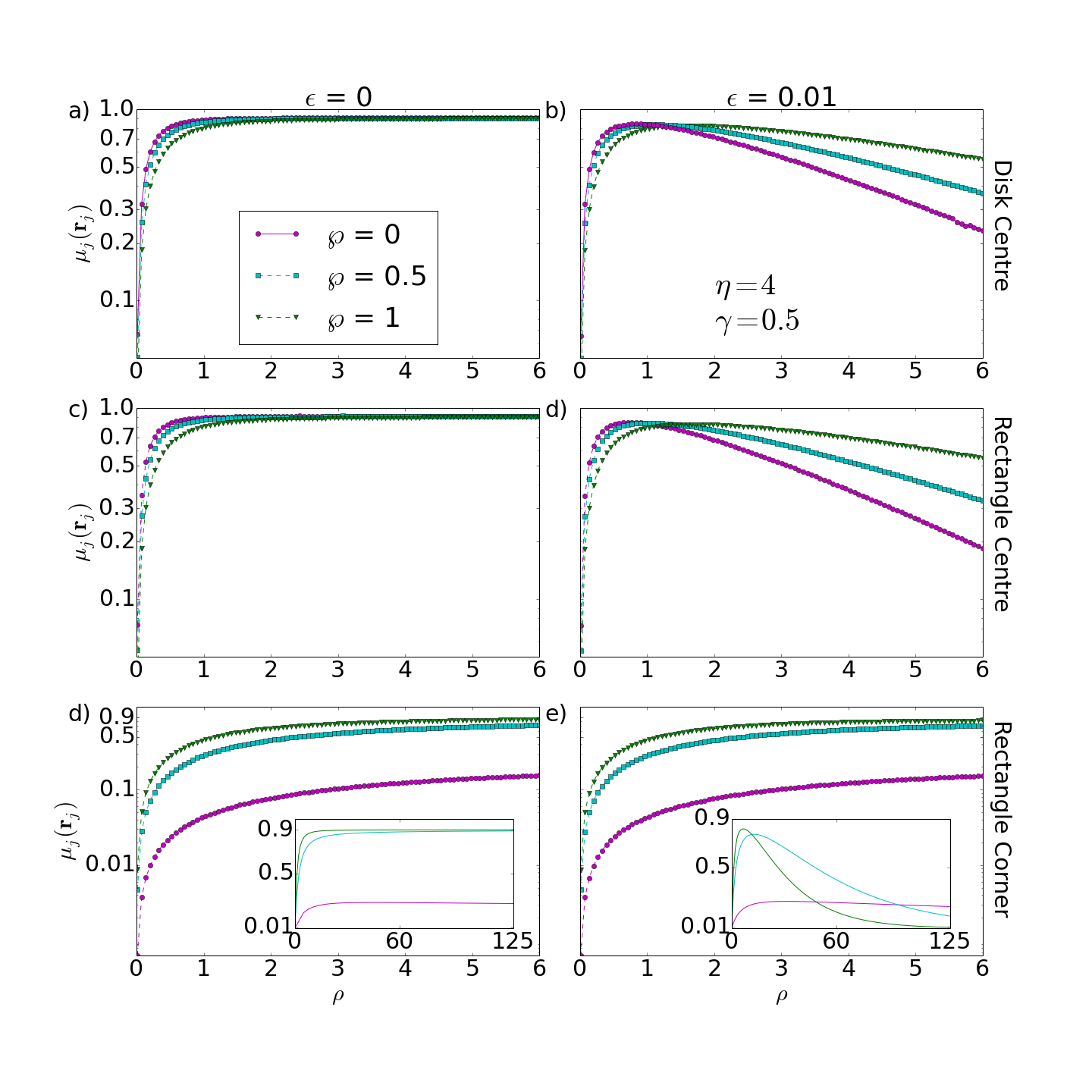}
\caption{A comparison between the spatial density $\mu(\br_j)$ for a receiver located different positions within a circular and square domain where we use numerical integration of equation \eqref{eq:mu}, using \eqref{eq:SINR_disk_eta_4}, where $\br_j \in\V $ are the Cartesian coordinates of the receiver.  In the above figures  $\epsilon = 0$  in the left panel and $\epsilon = 0.01$ in the right and we take $R = 5$, $a = b = \frac{\sqrt{\pi}R}{2}$ such that the two domains have the same volume V.  We use the model where the network is operating in a cluttered environment,  $\eta = 4$,  and not all nodes are interfering with each other, $\gamma = 1/2$.}
\label{fig:fig5}
\end{figure}
Closed form calculation of \eqref{eq:mu} is possible for the SNR case where $\gamma=0$ and $\eta=2$ giving 
\es{\mu(\boldsymbol{r}_j ) &= (N-1)\frac{\wp_p\pi }{4V}\biggl\{\erf\left(a-x_0\right) + \erf\left(a+x_0\right)\biggl\}\\
&\times\biggl\{\erf\left(b-y_0\right)+ \erf\left(b+y_0\right)\biggl\}\\
&+   (N-1)\frac{9(1-\wp_p)e^{-2(ax_0 + by_0)}}{256(ab)^3} \biggl\{e^{\left(x_0 - a\right)^2}\\
&\biggl[-2(a+x_0)e^{2ax_0} + \sqrt{\pi}(2x_0^2 - 2a^2)\erf\left(a-x_0\right)\biggl] \\
&+ e^{-\left({a^2 + x_0^2}\right)}\biggl[-2(a-x_0) \\
&+ \sqrt{\pi}(2x_0^2- 2a^2)\erf\left({a+x_0}\right)e^{\left({a+x_0}\right)^2}\biggl]\biggl\}\\
&\biggl\{e^{\left(y_0 - b\right)^2}\big[-2(b+y_0)e^{2by_0}\\
&+ \sqrt{\pi}( 2y_0^2- 2b^2)\erf\left(b-y_0\right)\biggl]\\
&+ e^{-\left(b^2 + y_0^2\right)}\biggl[-2(b-y_0) \\
&+ \sqrt{\pi}( 2y_0^2- 2b^2)\erf\left({b+y_0}\right)e^{\left({b+y_0}\right)^2}\biggl]\biggl\}
\label{eq:mu_SNR}
}
It should be noted that a closed form expression can be achieved for the unit disk model where $\gamma = 0$ and $\eta \to \infty$ but we only give here explicit expressions for the mean degree at the centre, corner and the midpoint of the edges for the sake of brevity. 
\es{
\mu(0,0) &= \frac{(N-1)\wp_p\pi}{\overline{l}\mathcal{V}} + \frac{(N-1)(1-\wp_p)\pi}{42(ab)^3} \big\{\\&- 6(a^2 + b^2)+ 24(ab)^2 \big\}\\
\mu(a,0) &= \frac{(N-1)\wp_p\pi}{8 \overline{l}(ab)} + \frac{(N-1)(1-\wp_p)}{426(ab)^3}\big\{5\pi \\&- 64a - 30\pi b^2 + 320ab^2\big\}\\
\mu(0,b) &= \frac{(N-1)\wp_p\pi}{8 \overline{l}(ab)} +\frac{(N-1)(1-\wp_p)}{426(ab)^3}\big\{5\pi \\&- 64b - 30\pi a^2  + 320a^2b\big\}\\
\mu(a,b) &= \frac{(N-1)\wp_p\pi}{16 \overline{l}(ab)} + (N-1)\frac{(1-\wp_p)}{854(ab)^3}\big\{5\pi \\&- 64(a+b) + 120ab \big\}
\label{eq:mu_disk}
}

From the closed form expressions of \eqref{eq:mu_SNR}, \eqref{eq:mu_disk}  we can clearly see that the number of successful transmissions in the SNR and unit disk model $\mu$ grows linearly with $\rho=N/V$. 
We note that an analytical expression for \eqref{eq:mu} cannot be given for the SINR case $\gamma > 0$. 

In contrast to the case of $\gamma = 0$, for interference limited networks (i.e. $\gamma \neq 0$) for a circular domain, the spatial density of successful transmissions $\mu$ plateaus for $\epsilon=0$ and is unimodal (i.e. has a single maximum) for $\epsilon>0$, indicating a deterioration in performance at large densities, Fig.~\ref{fig:fig5}\CB{a}\Ck{}. 
This suggests that $\epsilon \neq 0$ may give a more accurate representation of what happens in real life networks. 
Intuitively we can see this as if we start with a single node in our system and continue to add more then the amount of successful transmissions that can be achieved will initially increase since at low densities the effect of the interference field will not be as great. 
However, at a particular density we would expect this number to stop increasing and start decreasing since the strength of the interference will have increased such that it restricts the number of successful transmissions.

Both the uniform and RWPM distribution have a low $\mu$ at the centre but we see much faster deterioration in performance as the density increases  for $\wp = 0$, a result of the spatial distribution of nodes; for a network with roughly 235 nodes, $\mu_{(0,0)} \approx 0.5$. 
We note that Fig.~\ref{fig:fig5}\CB{a}\Ck{} and Fig.~\ref{fig:fig5}\CB{c}\Ck{} are very similar due to the chosen parameters, likewise for Fig.~\ref{fig:fig5}\CB{b}\Ck{} and Fig.~\ref{fig:fig5}\CB{d}\Ck{}.

Both Fig.~\ref{fig:fig5}\CB{d,5e} \Ck{} show that for $\wp = 0$ the behaviour of $\mu$ near the borders is very poor. Even though a receiver at the border (i.e. at the corner) is expected to have a higher connection probability $H_{ij}$ in an interference limited network, the spatial density of successful transmissions suffers greatly when the density of nodes is small. 
Interestingly, we see that even for $\epsilon = 0.01$ the behaviour of $\wp = 0$ changes very little as $\rho$ increases, insert of Fig.~\ref{fig:fig5}\CB{e}\Ck{}, and in fact only as the number of nodes in the network becomes very large, $\approx 10^3$, do we begin to see more noticeable degradation in $\mu$, highlighting the non-trivial trade-off between interference and boundary effects.
Interestingly when the distribution of nodes is a mixture of RWPM and Uniform, $\wp = 0.5$, we see that the performance at the borders is significantly improved compared with that of $\wp = 0$.

In terms of applications of these results, under the RWPM it suggests that areas such as the centre of cities where the largest number of mobile devices are typically found, a drop in ad hoc network performance is expected due to interference, results which are consistent with that of \cite{gong2014interference}. 
This is further intensified near the borders simply due to the low spatial density of transmitting nodes; however insert of Fig.~\ref{fig:fig5}\CB{e}\Ck{} indicates that the network performance could be greatly improved by having a hybrid network.
Efficient, interference and mobility aware MAC protocols would have to be  developed in line with the above results as to optimise channel access and thus connectivity throughout the network domain $\V$.

The calculations shown in  Fig.~\ref{fig:fig3},~\ref{fig:fig3b},~\ref{fig:fig5} were done using the approximate pdf's as the exact ones would not change the result qualitatively.

\section{Discussion and Conclusions \label{sec:Conclusion}}

The random waypoint model assumes nodes moving from waypoint to waypoint in a random fashion.
The resulting spatial distribution of nodes is not uniform but rather is concentrated in the domain's bulk.
As such, when wireless nodes access the common Hertzian medium without any collision avoidance mechanism (or similar) signals will interfere with each other at the receiver end causing sever packet losses which need to be catered for through retransmissions causing further delay and requiring additional signalling overheads.

In this paper we have explored the Random Waypoint Mobility Model where we have given an exact expression for the spatial distribution of nodes in a rectangle and  compared this with earlier approximations. A closed form expression for the connection probability for a receiver positioned at the centre of a circular domain was also given, and numerical calculations for receivers positioned at different locations and in different domains. Due to the high number of nodes in the bulk for the RWPM, it was shown that a successful connection was less likely to happen since the interference is greater, where the converse was true at the edges; an effect that was magnified in the rectangular region due to its corners. An exact expression was calculated for the mean degree of a noisy network in free space and shown to grow linearly with density. 
Numerical calculations were used to express the spatial density of successful transmissions in different parts of the domain where the performance of a network locally is characterised by the trade off between the density of nodes and the resulting interference field (i.e. too few nodes limits the number of possible connections, too many results in too much interference). 
Finally, we have illustrated that the non-singular path loss function gives a more realistic analysis of the networks performance and shows how important metrics like the mean degree deteriorates at higher densities.

These results highlight one important challenge facing any future deployment of MANETs, namely how can the interference experienced from neighbouring nodes be mitigated if the underlying network intensity follows an inhomogeneous spatial distribution. 
We see that in the RWPM nodes are focused predominately in the bulk of the domain and thus more mobile devices are competing for the same  amount of limited resources so efficient protocols would be needed to ensure fair channel access.
This observation is expected to hold under more advanced protocols (e.g. CSMA/CA) than the simple ALOHA assumed herein since despite careful interference avoidance mechanisms, different parts of the network have different densities of transmitters and receivers which therefore suffer from different levels of co-channel interference. 
Throughout this paper we do not assume any neighbour association mechanism which is typical in cellular networks and as such could be a line of future investigation.
As a final remark, we also point towards the common assumption that the most efficient way to transmit data between nodes in a multi-hop fashion is via the shortest path.  However, with an inhomogeneous distribution and  interference limited environment,  this may not be the most effective since going through areas of high density, the levels of  interference increase so data may have to be continually resent thus increasing the transmission time between sender and receiver. Instead the optimal solution might involve minimising both the path and the interference.

\section*{Acknowledgments}
The authors would like to thank the directors of the Toshiba Telecommunications Research Laboratory for their support.
This work was supported by the EPSRC [grant number EB/N002458/1].
In addition, Pete Pratt is partially supported by an ESPRC Doctoral Training Account.


\bibliographystyle{ieeetr}
\bibliography{ref_bib}

\end{document}